\documentclass[aps,prl,twocolumn,unsortedaddress]{revtex4-2}
\usepackage{amsmath}
\usepackage{amsfonts}
\usepackage{amssymb}
\usepackage{graphicx}
\usepackage[dvipsnames]{xcolor}
\usepackage{float}
\usepackage{siunitx}
\usepackage[version=4]{mhchem}
\usepackage{hyperref}

\usepackage{xcite}
\usepackage{xr}
\makeatletter
\newcommand*{\addFileDependency}[1]{
  \typeout{(#1)}
  \@addtofilelist{#1}
  \IfFileExists{#1}{}{\typeout{No file #1.}}
}
\makeatother

\DeclareSIUnit[number-unit-product = {\,}] \cal{cal}

\newcommand{\kB}{k_\mathrm{B}}

\newcommand{\lD}{\lambda_\mathrm{D}}
\newcommand\dv{\,{:}\,}

\newcommand{\dep}[2]{\ensuremath{\frac{\partial #1}{\partial #2}}}
\newcommand{\depn}[3]{\ensuremath{\frac{\partial^{#3} #1}{\partial #2^{#3}}}}

\newcommand{\appropto}{\mathrel{\vcenter{
  \offinterlineskip\halign{\hfil$##$\cr
    \propto\cr\noalign{\kern2pt}\sim\cr\noalign{\kern-2pt}}}}}

\newcommand{\BR}[1]{{\color{black}#1}}

\newcommand{\IP}[1]{{\color{black}#1}}

\usepackage{ulem}

\begin{document}
	\title{Charging dynamics of electric double layer nanocapacitors in mean-field}
	\author{Ivan Palaia}
    \affiliation{\BR{Department of Physics, King’s College London, WC2R 2LS, United Kingdom}}
    \affiliation{Institute of Science and Technology Austria, 3400 Klosterneuburg, Austria}
    \author{Adelchi J. Asta}
    \affiliation{\IP{Dessia, 92160 Antony, France}}
    \author{\BR{Megh Dutta}}
	\affiliation{\IP{Sorbonne Université, CNRS, Physico-chimie des Électrolytes et Nanosystèmes Interfaciaux, PHENIX, 75005 Paris, France}}
    \author{Patrick B. Warren}
	\affiliation{Hartree Centre, Science and Technology Facilities Council (STFC), Sci-Tech Daresbury, Warrington WA4 4AD, United Kingdom}
    \author{Benjamin Rotenberg}
	\affiliation{Sorbonne Université, CNRS, Physico-chimie des Électrolytes et Nanosystèmes Interfaciaux, PHENIX, 75005 Paris, France}
	\affiliation{Réseau sur le Stockage Electrochimique de l’Energie (RS2E), FR CNRS 3459, 80039 Amiens Cedex, France}
	\author{Emmanuel Trizac}
	\affiliation{Universit\'e Paris-Saclay, CNRS, LPTMS, 91405 Orsay, France}
	\date{\today}

 \begin{abstract}
    An electric double layer capacitor (EDLC) stores energy by modulating the spatial distribution of ions in the electrolytic solution that it contains. We determine the mean-field time scales for planar EDLC relaxation to equilibrium, after a potential difference is applied. We tackle first the fully symmetric case, where positive and negative ionic species have same valence and diffusivity, and then the general, more complex, asymmetric case. Depending on applied voltage and salt concentration, different regimes appear, revealing a remarkably rich phenomenology relevant for nanocapacitors.
 \end{abstract}

	\maketitle


Two conductive surfaces separated by an ionic solution form an electric double layer capacitor (EDLC), that stores electrostatic energy by modulating the distribution of charged species in solution~\cite{salanne_efficient_2016,simon_perspectives_2020}. Nanoporous conductive materials offer an optimized contact between the electrolyte and the electrode, where charge storage occurs, leading to specific capacities as large as \SI{100}{\farad}/g of material \cite{Merlet2012}. Their ability to store and release charge much faster than in batteries, which involve electrochemical reactions~\cite{Sherrill2011}, allows their use in applications requiring high powers, from the recovery of breaking energy to electrical public transportation means covering short distances and recharging during stops~\cite{Raghavendra2020}. A promising use of so-called supercapacitors concerns the extraction of ``blue energy" from fresh and salty water, or conversely the desalination of water, using cycles of charge and discharge of capacitors~\cite{Brogioli2009, Boon2011, Janssen2014,suss_water_2015,simoncelli_blue_2018}. Finally, it is now possible to use electrodes in nanocapacitors and nanofluidic devices to study extremely small volumes of electrolytes~\cite{sun_electrochemistry_2008,mathwig_electrical_2012,rassaei_hydrodynamic_2012}.

Predicting the charging dynamics of EDLCs is essential, because tuning the related characteristic time \IP{can} maximize efficiency. \IP{For instance, understanding the charging dynamics is critical to identifying optimal charging protocols that may minimize (dis)charging times, while meeting technological constraints \cite{Breitsprecher2018,Breitsprecher2020,phd}.} For both aqueous solutions and ionic liquids, the effects of ion correlations and size on the charging dynamics have been described by density functional theories \cite{Kornyshev2007, Jiang2014, Goodwin2017, Ma2022}, lattice models \cite{Lee2015} and molecular simulations \cite{Limmer2013, Kondrat2014, Pean2014, Breitsprecher2018, Noh2019, Lian2020, scalfi_molecular_2021, HoangNgocMinh2022, Jeanmairet2022}. More fundamental studies rely on mean-field continuum models of the electrolyte \cite{Bazant2004, Beunis2008,Janssen2018}, in planar and non-planar geometries \cite{Janssen2019, Yang2022, Werkhoven2018}, which can also be simulated using lattice-based models to capture electrokinetic couplings~\cite{asta_lattice_2019}. In this context, particularly well studied is the linear regime (small applied voltage) for a symmetric electrolyte (cations and anions with same valence and diffusivity) \cite{Bazant2004,Beunis2008,Janssen2018}. 
\IP{While the large-voltage regime has received attention for symmetric systems \cite{Bazant2004, Beunis2008}, its general understanding remains incomplete. The asymmetric electrolyte case, especially at large voltages, is unexplored, despite ion asymmetry being the rule rather than the exception. Previous studies have only considered a 1:1 electrolyte with unequal diffusivities at low voltage \cite{Balu2018}. Furthermore, non-linear effects arising from the fact that the number of ions enclosed in real capacitors is fixed, rather than their chemical potentials, are scarcely characterized, with insights only available for the symmetric case \cite{Bazant2004, Beunis2008}.}
In the Letter, we fill these gaps and identify how the time scales governing relaxation to equilibrium depend on the key parameters of the system: applied voltage, ion concentration and distance between electrodes. We find an unexpectedly rich regime diagram, highlighting the importance of finite system size and ion asymmetry.  

\begin{figure}
	\centering
	\includegraphics[width=0.8\columnwidth]{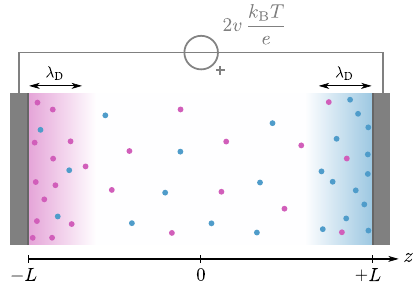}
	\caption{An ideal electric double layer capacitor (EDLC).
 The total amount of salt is fixed and, in the linear regime, it defines the thickness of the double layer $\lD$. 
	}
	\label{figm:system}
\end{figure}

Our model system is represented in Fig.~\ref{figm:system}. The plates, distant $2L$ from each other, are ideal conductors whose dielectric mismatch with the solution we neglect. To allow analytical and computational treatment, water is supposed to have constant permittivity $\varepsilon_0\varepsilon_r$ and correlations between ions are neglected -- a condition known as weak coupling \cite{Naji2013}. The system is at equilibrium at zero potential difference ($2v=0$) for times $t<0$: the two ionic species, positive and negative, are homogeneously distributed, the solution is locally neutral everywhere and the plates are uncharged. At time $t = 0$ we instantaneously switch on the potential $2v > 0$, that we measure in units of thermal energy per elementary charge $\kB T/e$. The plates charge up and ionic concentration profiles change in response to this, obeying the Poisson-Nernst-Planck theory \cite{Hunter}. For two ionic species of valences $\pm q_\pm$ and diffusion coefficients $D_\pm$, this relates the electric potential $\phi(z,t)$ to the ion densities $n_\pm(z,t)$:
\begin{align}
\dep{n_\pm}{t}&=D_\pm \dep{}{z} \left(\, \pm \frac{q_\pm e}{\kB T} n_\pm\, \dep{\phi}{z} + \dep{n_\pm}{z}\right) \label{eqm:NP}
\\
-\depn{\phi}{z}{2}&=\frac{\rho}{\varepsilon_0\varepsilon_r}\,. \label{eqm:Poisson}
\end{align}
Here, 
$\rho(z,t)=q_+en_+(z,t)-q_-en_-(z,t)$ is the electric charge density. Eq.~\eqref{eqm:NP} is a continuity equation, whose current has a drift and a diffusion term. Its equilibrium zero-current state retrieves the Poisson-Boltzmann distribution \cite{Hunter}. Eq.~\eqref{eqm:Poisson} is the Poisson equation. 

We solve numerically Eqs.~\eqref{eqm:NP}-\eqref{eqm:Poisson} via a flux-conservative finite-difference integration scheme, described \IP{in our companion paper~\cite{supp}, which also includes a nondimensionalisation of the equations and a more detailed analysis of the regimes and times scales discussed below.} Boundary conditions are the desired potential difference and the vanishing of the ionic current at the electrodes. We work in the canonical ensemble, with no salt reservoir. We collect data for a range of applied voltages and salt concentrations spanning, respectively, 5 and 10 orders of magnitude. We quantify the initial densities $n^0_\pm=n_\pm(z,0)=\int_{-L}^L n_\pm(z,t) \,\mathrm{d}z / (2L)$ through the dimensionless quantity $\lD/L$, where $\lD=[ (q_+^2 n^0_+ + q_-^2 n^0_-) e^2 / (\kB T \varepsilon_0 \varepsilon_r) ]^{-1/2}$ and defines the Debye screening length. The salt concentration is $n^0=n^0_+/q_+=n^0_-/q_-$. We analyse the time evolution of the surface charge density $\sigma(t) \propto  \partial \phi / \partial z (\pm L,t) $ (equal and opposite on the two electrodes, as the current entering the generator equals the exiting one),
and of the ionic densities $n_\pm(z,t)$ through the proxy $\rho(\pm L, t)$.
We characterise the relaxation to the equilibrium values $\sigma_\mathrm{eq}=\sigma(t\to\infty)$ and $\rho_\mathrm{eq}^\pm=\rho(\pm L, t\to\infty)$ by inspecting all curves for exponential or linear relaxation rates.

\begin{figure}
	\centering
	\includegraphics[width=1\columnwidth]{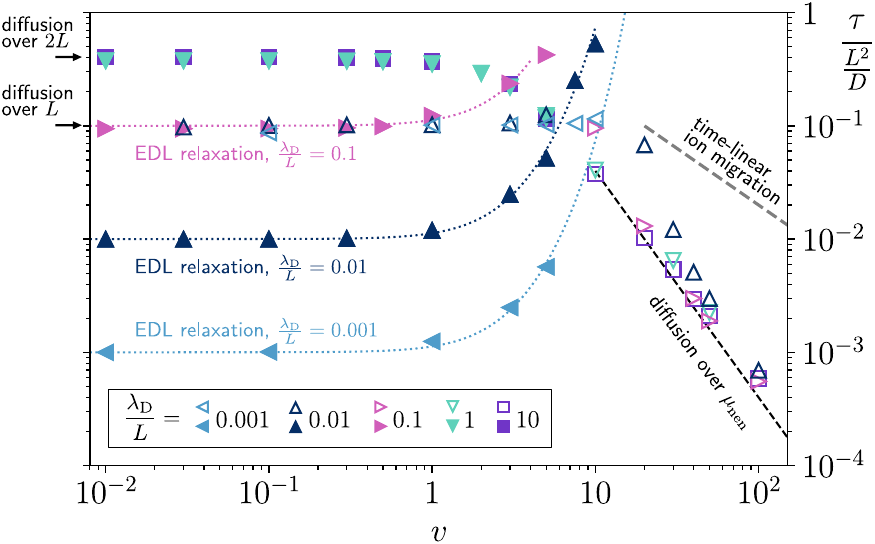}
	\caption{Exponential relaxation times $\tau$ extracted from linear fits of $\log(\sigma(t))$ vs $t$, as a function of dimensionless voltage $v$. 
	For given $\lD$ and $v$, two different relaxation processes are often seen in $\sigma(t)$ (see Fig.~\ref{figm:phasesym}): filled  symbols indicate the early-time process, whereas empty symbol the late-time process, when present. For $v\le1$ and $\lD/L\ge1$, the relaxation is purely diffusive and takes place on a scale $2L$. For $v\le1$ and $\lD/L\ll1$, the double layer relaxes at early times on a time ${L\lD}/{D}$, that extends into the nonlinear regime as $({L\lD}/{D})\cosh({v}/{2})$ (dotted curves). This is followed by a slower diffusive relaxation over a length $L$, signaling depletion (empty symbols). For $v\gg1$, collective ion migration causes full depletion: this early-time process is not shown here because it is non-exponential (the gray dashed line however shows its time scale for the unscreened regime of Fig.~\ref{figm:phasesym}, where the process is linear). At late times, a fast diffusive relaxation follows (empty symbols), signaling ion rearrangement inside counterionic double layers of thickness $\mu_\mathrm{nen}$ (the Gouy-Chapman length).
	}
	\label{figm:tausym}
\end{figure}

\begin{figure*}
	\centering
	\includegraphics[width=0.95\textwidth]{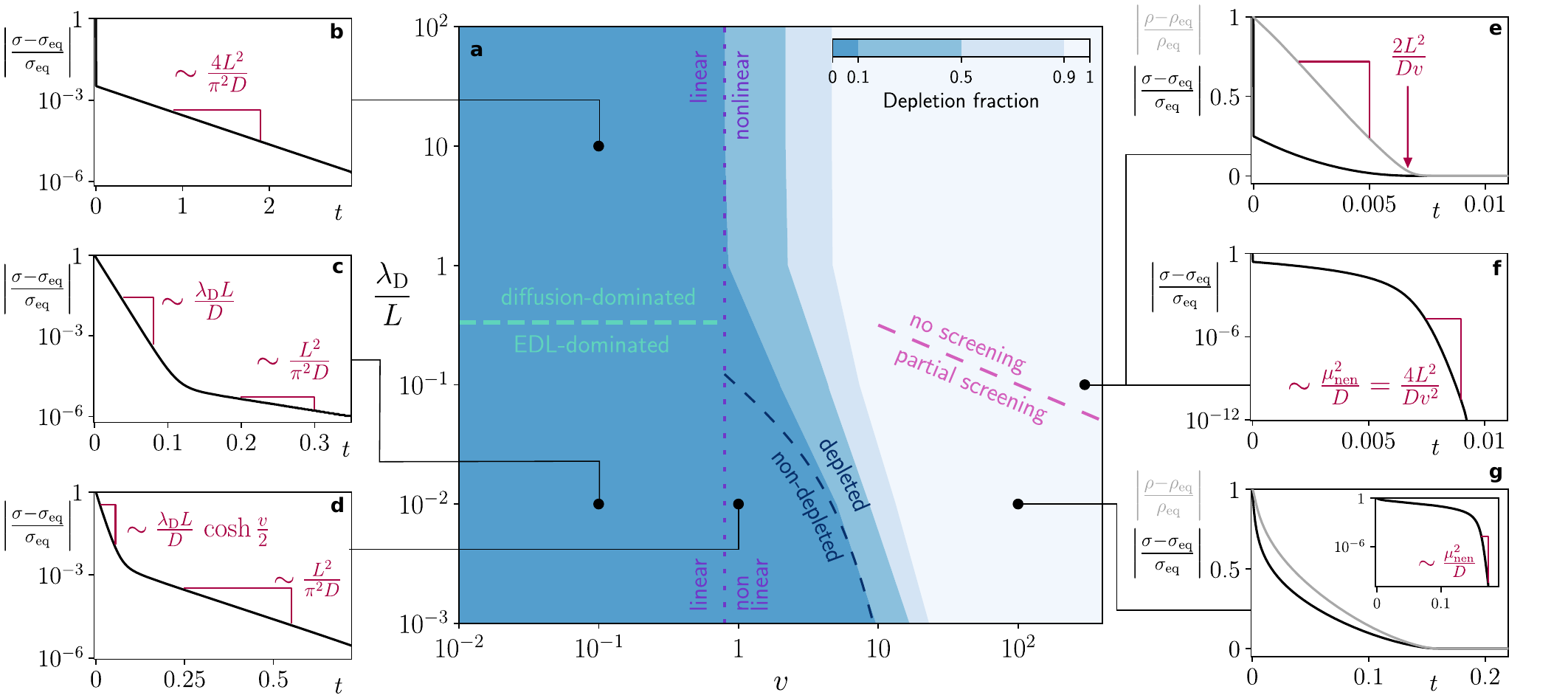}
	\caption{a) Regime diagram for the symmetric electrolyte case ($D_+=D_-=D$, $q_+=q_-=1$). Five different regimes are separated by the boundary lines discussed in the text. b-g) In black, the relative difference between the instantaneous electrode charge density $\sigma(t)$ and its equilibrium value $\sigma_\mathrm{eq}=\sigma(t\to \infty)$. In addition, in e and g, the gray curves show the relative difference between volume charge density $\rho(t)$ at contact with the electrode and its equilibrium value $\rho_\mathrm{eq}$. 
    Time $t$ has units of $L^2/D$. In crimson, scaling of relaxation times as extracted from linear fits and confirmed analytically \cite{supp}.}
	\label{figm:phasesym}
\end{figure*}

For the symmetric electrolyte case ($D_+=D_-=D$ and $q_+=q_-$), relaxation times $\tau$ are summarised in Fig.~\ref{figm:tausym} and are used to formulate the regime diagram of Fig.~\ref{figm:phasesym}a. For small $v$, the system is treatable analytically \cite{Bazant2004, Janssen2018} and presents an infinite series of exponential relaxation times. For $\lD/L>1$, they all scale as $\sim L^2/D$, the largest being $4L^2/(\pi^2 D)$: this is the dominant time scale and the only one clearly visible (Fig.~\ref{figm:phasesym}b) and it is a signature of ions diffusing over a length $2L$ toward the oppositely charged plate. For $\lD/L<1$, subdominant times are of order $\lD^2/D$ while the dominant one scales as $L\lD/D$ (Fig.~\ref{figm:phasesym}c). The exact expression of relaxation times in the linear regime was obtained by \cite{Janssen2018} and we present an alternative derivation in \cite{supp}. Interestingly, already from the linear regime, at small $\lD/L$, a sign of nonlinearity appears, termed depletion: due to asymmetric accumulation in the electric double layers (EDL), the salt concentration decreases in the middle of the capacitor. Indeed, at equilibrium, for each species, the EDL next to the oppositely charged electrode is more populated than the other EDL is depopulated. After most of the EDL has built up, a neutral excess of ions remains around $z=0$ and diffuses away in a time $L^2/(\pi^2 D)\simeq10^{-1}L^2/D$. This diffusion, occurring on a length $L$, is visible in Fig.~\ref{figm:tausym} ($v<1$, empty symbols, representing late-time relaxation) and in the long-time slope of Fig.~\ref{figm:phasesym}c. \IP{We note that some electrode charge appears instantly at switch-on — that of an ideal capacitor under voltage in a neutral homogeneous dielectric. At low concentrations ($\lD/L > 1$), this accounts for most of the equilibrium charge (hence the jump at $t=0^+$), whereas for $\lD/L>1$, ions dramatically increase energy storage and this initial charge is negligible.}

In Fig.~\ref{figm:phasesym}a, $v>1$ defines the nonlinear regime. Depletion, quantified through the depletion fraction $n_\pm(0,t\to\infty)/n_\pm^0$ and indicated by shadings of blue, dramatically affects this regime. However, at small $\lD/L$ the ion concentration is large enough to make depletion a second order effect and a non-depleted, purely nonlinear regime is visible. Such a window is delimited by the condition $2\,(\lD/L)\, \sinh(v/2) \ll1$ (dark-blue dashed curve), matching the numerical calculations. This is obtained through the Grahame equation, relating potential and electrode charge \cite{Hunter,Andelman2010}. We impose that the latter is much less than the charge of all oppositely charged ions in the system: $\sigma_\mathrm{eq}\ll 2n^0 L$ (see \cite{supp}). Relaxation, in Fig.~\ref{figm:phasesym}d, is governed at short times by a new exponential time scale for EDL formation, increasing with $v$ as shown in Fig.~\ref{figm:tausym} (symbols on dotted line). A RC-circuit equivalence shows that this time scale is $(\lD L/D)\cosh(v/2)$ (dotted lines) and reflects the increased charge, i.e.~capacitance, of the nonlinear EDL. At late times, as in the linear regime, depletion manifests as a diffusive relaxation rate at late times, corresponding to the relaxation of the neutral excess of ions. 

As $v$ increases further into the nonlinear regime, the system rapidly transitions to a fully depleted final state (depletion fraction $\simeq1$, i.e.~practically no ions in the bulk) and the physics changes drastically. To understand it, we focus first on the top-right part of Fig.~\ref{figm:phasesym}a. For such strong voltages, ionic screening has a relatively small effect on the electrode charge, so that ions are pulled at constant velocity toward the oppositely charged wall. Charge density in the EDL grows linearly in time, as shown in Fig.~\ref{figm:phasesym}e (gray). Also, since the applied voltage results from the sum of the electric fields due to the electrodes' charge and to the ionic charge, the latter are linearly related. Because the non-neutral bulk portion increases linearly with time, the electric field due to the solution is parabolic in time, and so is $\sigma$ (black)\BR{, as already reported for the fully symmetric case in~\cite{Beunis2008}}. Once the two purely counterionic EDLs are formed, a final exponential relaxation occurs (Fig.~\ref{figm:phasesym}f)\BR{, not described in~\cite{Beunis2008}}. The only meaningful length scale is now the Gouy-Chapman length, the distance over which an isolated counterion can be dragged away from the electrode, with an energy budget $\kB T$ \cite{Hunter, Andelman2010}; it sets the extension of the double-layer in a salt-free regime. Since one electrode carries a larger charge than its counterions, this half-system is not electroneutral and the right Gouy-Chapman length reads $\mu_\mathrm{nen}=2\varepsilon_0\varepsilon_r\kB T/(e^2 q \sigma_\mathrm{res})$, where `nen' stands for non-electroneutral and $\sigma_\mathrm{res}=\sigma_\mathrm{eq}-2n^0L$ is the part of electrode charge not neutralised by counterions \cite{supp}. As $\sigma_\mathrm{res}\appropto v$, the late relaxation time is $\mu_\mathrm{nen}^2/D\propto v^{-2}$ (dashed black line in Fig.~\ref{figm:tausym}).  

We have just assumed that the ion dynamics do not perturb much the externally imposed electric field in the EDL (yet it sensibly affects electrode charge relaxation, as seen in Fig.~\ref{figm:phasesym}e). This only occurs if $\sigma_\mathrm{eq}\appropto v$ is $\gg 2n^0L$, which defines the region above the pink line in Fig.~\ref{figm:phasesym}a. Below such line, ions partially screen the applied field, by an amount that varies during relaxation. In this regime the early-time relaxation dynamics is neither linear, nor quadratic or exponential (Fig.~\ref{figm:phasesym}g). However, it ends as abruptly as in the unscreened regime and, at late times, the purely counterionic EDL relaxes exponentially over the Gouy-Chapman length $\mu_\mathrm{nen}$ (inset).

\begin{figure*}
	\centering
	\includegraphics[width=\textwidth]{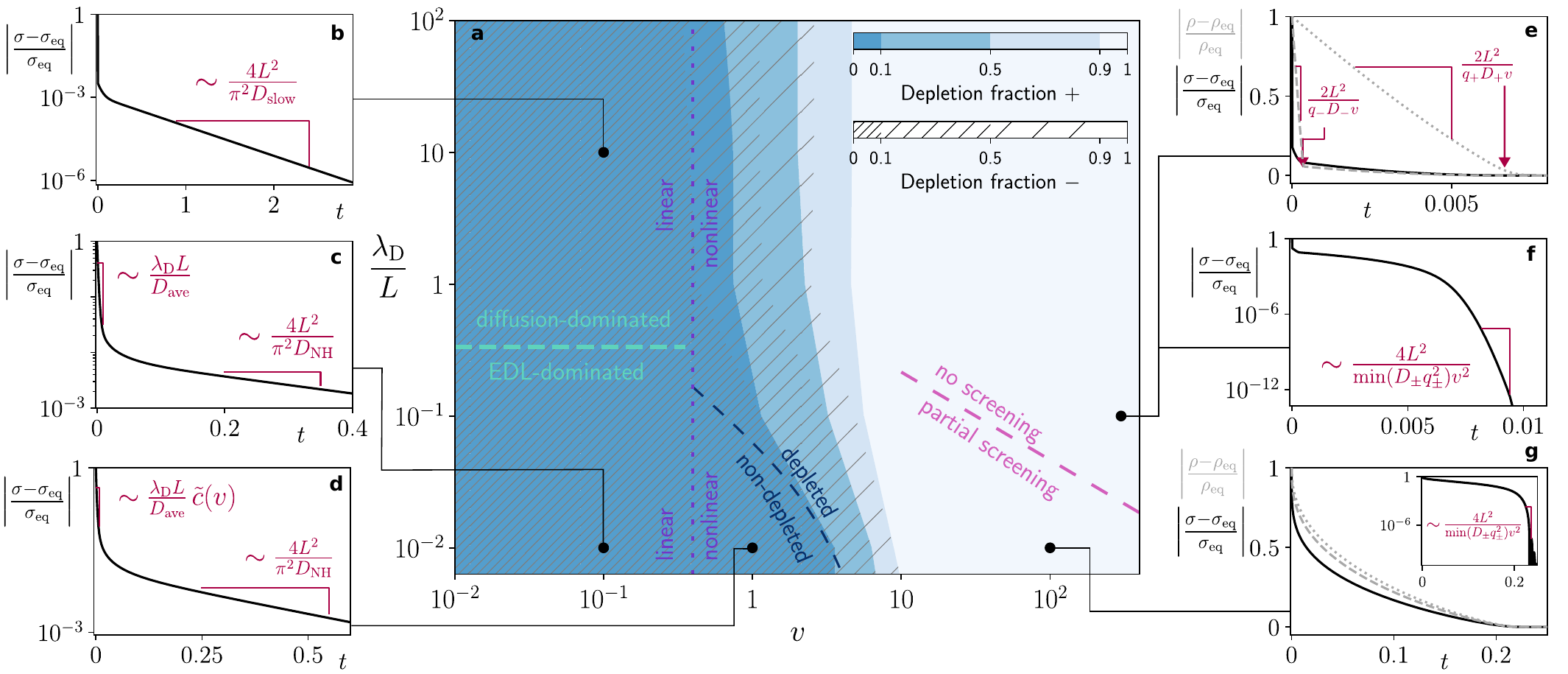}
	\caption{a) Regime diagram for the completely asymmetric case ($D_+/D_-=1/10$, $q_+=1, q_-=2$). Depletion of positive ions (blue tones) is distinct from that of negative, more charged, ions (hatch patterns). b-g) As in Fig.~\ref{figm:phasesym}. In e and g, dotted and dashed grey lines represent the charge densities at the negative and the positive electrodes, respectively; in the symmetric case, these were equal.  Time $t$ has units of $L^2/D_+$.
	} 
	\label{figm:phaseasym}
\end{figure*}

The general asymmetric case is described in Fig.~\ref{figm:phaseasym}. To obtain clear separations between the relevant time scales, we choose $D_+/D_- = 1/10$, $q_+ \dv q_- = 1 \dv 2$, and call positive ions slow and negative ions fast;  other choices lead to analogous results \cite{supp}.
Several differences arise compared to the symmetric case described so far, making the phenomenology even richer. In the linear, large $\lD/L$ regime, the two ionic species are completely decoupled. Each of the infinitely many relaxation modes from the symmetric case splits into two, so that half of them are proportional to $L^2/D_+$ and half to $L^2/D_-$; the slowest mode concerns the diffusion of the slowest species over a length $2L$, so its characteristic time is $4L^2/(\pi^2 D_\mathrm{slow})$, with $D_\mathrm{slow}=D_+$ (Fig.~\ref{figm:phaseasym}b). Valences do not affect relaxation times, but they play a role in determining the weight of each mode \cite{supp}. 

In constrast, at small $\lD/L$ (Fig.~\ref{figm:phaseasym}c) valences matter: as the bulk conductivities due to the two species are additive and each of them is proportional to $q_\pm D_\pm$, a simple RC-circuit analogy shows that the relevant diffusivity is $D_\mathrm{ave}=(q_+ D_+ + q_- D_-)/(q_+ + q_-)$, and the EDL forms on a timescale $\lD L/D_\mathrm{ave}$ \cite{supp}. At late times, relaxation is due to the neutral excess diffusion governed instead by the Nernst-Hartley diffusivity $D_\mathrm{NH}=(q_+ + q_-) D_+ D_- / (q_+ D_+ + q_- D_-)$ \cite{Robinson1959}, relevant also in the context of impedance spectroscopy \cite{Barbero2005, Barbero2007, Barbero2008}. Indeed, in relaxing the neutral excess by diffusion, positive and negative ions must move together, with the slow species slowing down the fast and the fast pulling the slow: this is reflected by a friction (inverse diffusivity) which is the average of the frictions of the two species. In the ion-symmetric case, relaxation of the neutral excess occurs over a length $L$, from the exact centre to the electrodes, effectively forbidding a diffusive mode with wavelength $2L$; ionic asymmetry relaxes this constraint, so that the late-time characteristic time is now $4L^2/(\pi^2 D_\mathrm{NH})$. This neutral excess relaxation is a purely linear phenomenon and is not due to depletion (that still happens, but on a faster, hidden scale here). As the fast species tends to relax according to the instantaneous distribution assumed by the slow species, an overcrowding of carriers occurs at the electrode of same charge as the slow ions: these move away slowly, so that the fast ions, in the attempt to equilibrate the EDL, arrive in larger amounts than needed for equilibrum. Eventually, as the slow ions gradually move away, the overabundant fast ones also leave the EDL in the observed relaxation of the neutral excess by diffusion. 

Macroscopic depletion, at large $\lD/L$, now occurs at different voltages for the two species. This is shown by the mismatch between colour gradient and hatching patterns in Fig.~\ref{figm:phaseasym}a. The linear regime shrinks to the region $v < q_\mathrm{max}^{-1} = q_-^{-1}$. In the intermediate region $q_\mathrm{max}^{-1}<v<q_\mathrm{min}^{-1}$, sharply defined for strong valence asymmetry, the depletion mismatch results in a rapid relaxation of the higher-charge (depleted) species, followed by a linear-regime-like relaxation of the smaller-charge (non-depleted) species \cite{supp}. At small $\lD/L$, where nonlinear features emerge before bulk depletion, this effect disappears and the two species are equally depleted at given $v$. The analytic expression for the onset of depletion (dashed blue line) is given in \cite{supp}: as depletion is measured at equilibrium, it depends on $q_\pm$ but not on $D_\pm$.

The non-depleted nonlinear regime (Fig.~\ref{figm:phaseasym}d) confirms the importance of $D_\mathrm{ave}$ and $D_\mathrm{NH}$. In particular, during early-time EDL build-up, the $\cosh$ factor accounting for increased EDL capacitance in the symmetric case is replaced by $\tilde{c}(v)$, whose cumbersome expression we work out in \cite{supp}. For $v\gg 1$, asymmetry always enhances capacitance, thereby increasing the relaxation time compared to the $1\dv 1 $ case. 

Finally, in the fully depleted, unscreened regime (Fig.~\ref{figm:phaseasym}e), the situation is analogous to the symmetric case, but the two trains of positive and negative ions are now dragged by the electric field at different velocities. This results in $\rho$ relaxing with different time scales at the two electrodes (dashed and dotted gray). The electrode charge (black) is piecewise quadratic. At late times, the only visible relaxation is due to the counterions of the slower EDL, each with its own $\mu_{\mathrm{nen}\pm}$. The characteristic time is then the largest one between the times $\mu_{\mathrm{nen}\pm}^2/D_\pm = 4 L^2/(D_\pm q_\pm v^2)$, which in Fig.~\ref{figm:phaseasym}f is given by the slower and lower-charge species. Such relaxation process is also the last one to happen in the partially screened regime (Fig.~\ref{figm:phaseasym}g), while, again, early-time behavior is neither exponential nor linear.

In summary, we identified the dominant relaxation processes within mean-field and their often counter-intuitive characteristic times, for the whole parameter space. This work provides a long-missed, easy-access frame of comparison for all theories intended to incorporate steric effects and ionic correlations or aimed at describing ionic liquids. While some of these have pointed out the limits of mean-field \IP{(we discuss in~\cite{supp} the relevance of non-idealities in the various regimes)}, many have already confirmed its relevance for real physical systems \cite{Ma2022}. Understanding the relaxation dynamics paves the way to optimization strategies and to the design of supercapacitors.
\\

\textbf{Acknowledgements\ \ }
This work has received funding from the European Union's Horizon 2020 research and innovation program under the Marie Sk\l{}odowska-Curie grant agreements Nos. 674979-NANOTRANS and 101034413. This project received funding from the European Research Council under the European Union's Horizon 2020 research and innovation program (Grant Agreement No.~863473). \IP{This work has received funding from the European Union's Horizon Europe research and innovation program under the Marie Sk\l{}odowska-Curie grant agreement No. 101119598-FLUXIONIC.} B.R.~acknowledges financial support from the French Agence Nationale de la Recherche (ANR) under Grant No.~ANR-21-CE29-0021-02 (DIADEM).

\bibliography{library}

\end{document}